\documentclass[twoside]{article}
\usepackage{latexsym,fleqn,espcrc2}
\usepackage[dvips]{graphics}
\newcommand{\be}{\begin{equation}}
\newcommand{\ee}{\end{equation}}
\newcommand{\bea}{\begin{eqnarray}}
\newcommand{\eea}{\end{eqnarray}}
\newcommand{\bc}{\begin{center}}
\newcommand{\ec}{\end{center}}
\newcommand{\prd}{Phys.\ Rev.\ D }
\newcommand{\ovsix}{${\cal O}(v^6)$ }
\newcommand{\ovfour}{${\cal O}(v^4)$ }
\newcommand{\oas}{${\cal O}(\alpha_s)$ }

\begin{document}

\title{Unquenched Charmonium with NRQCD}
\author{Chris Stewart\address{Department of Physics \& Astronomy, York University,
Toronto, ON M3J 1P3, Canada.} and Roman
Koniuk\addressmark\thanks{This research was funded by the Natural
Sciences and Engineering Research Council of Canada.}}

\begin{abstract}
We present results from a series of NRQCD simulations of the
charmonium system, both in the quenched approximation and with
$n_{f} = 2$ dynamical quarks. The spectra show evidence for
quenching effects of $\sim 10 \%$ in the $S$- and $P$-hyperfine
splittings. We compare this with other systematic effects.
Improving the NRQCD evolution equation altered the $S$-hyperfine
by as much as $20$ MeV, and we estimate \oas radiative corrections
may be as large as $40 \%$.
\end{abstract}

\maketitle

\section{The current status of Charmonium}

Past lattice NRQCD simulations of charmonium have substantially
underestimated the hyperfine spin-splittings.  The introduction of
\ovsix corrections to the NRQCD Hamiltonian caused a dramatic
decrease in the S-hyperfine splitting \cite{TrottierQuarkonium},
which fell below $50 \%$ of the experimental value. These \ovsix
simulations also demonstrated a large dependence on the definition
of the tadpole correction factor. Even in the less-relativistic
$\Upsilon$ system, the same highly-improved NRQCD action has not
provided conclusive agreement with experiment
\cite{HighImpUps,UnqUps}.

Relativistic heavy quark actions have fared somewhat better,
though still significantly underestimate spin splittings. A tadpole- and
SW-improved charmonium simulation by UKQCD \cite{UKQCDRelCharm99}
gave the $S$-hyperfine splitting roughly $40 \%$ below experiment.
Unquenching should increase the hyperfine splittings---in
\cite{charm92}, quenching effects were projected to be as large as
$40\%$, though $5$--$15\%$ seems more likely
\cite{UnqUps,FermionLoops}.

We describe a series of NRQCD charmonium simulations with both
quenched and unquenched lattices. The effects of different tadpole
correction schemes are examined, and we find a rough estimate of
the effect of \oas radiative corrections. Finally, we note a
sizeable systematic effect due to an instability in the usual
heavy-quark evolution equation.

\section{The NRQCD and Gluon Actions}

The NRQCD quark propagator is found from an evolution equation,
\bea
\label{EvolutionEquation}
  G_{t+1} &=& \left ( 1-\frac{aH_{0}}{2s}
  \right )^{s} U_{4}^{\dag} \left (1-\frac{aH_{0}}{2s}\right)^{s}
  \nonumber \\
  & & \left (1-a \delta H \right) G_{t} \, ,
\eea
where the Hamiltonian $H = H_0 + \delta H$ is
\bea
  \lefteqn{H_0 = \frac{-{\bf \Delta}^{(2)}}{2M_{0}} \, , } \\
  \lefteqn{\delta H_{v^{4}} =
  \frac{-c_{1}}{8M^{3}_{0}} \left({\bf \Delta}^{(2)}\right
  )^{2}
  + \frac{ic_{2}g}{8 M^{2}_{0}} \left ( \tilde{{\bf \Delta}} \cdot
  \tilde{{\bf E}} - \tilde{{\bf E}} \cdot \tilde{{\bf \Delta}}
  \right ) \nonumber }\\
  \lefteqn{+  \frac{c_{3}g}{8 M^{2}_{0}} \sigma \cdot \left ( \tilde{{\bf
  \Delta}} \times \tilde{{\bf E}} - \tilde{{\bf E}} \times
  \tilde{{\bf \Delta}} \right )
  - \frac{c_{4}g}{2M_{0}} \sigma \cdot \tilde{{\bf B}} \nonumber }\\
  \lefteqn{+ \frac{c_{5}a^{2}}{24 M_{0}} {\bf \Delta}^{(4)}
  - \frac{c_{6}a}{16s M_{0}^{2}} \left({\bf \Delta}^{(2)}\right )^{2}
  \, ,}\\
  \lefteqn{\delta H_{v^{6}} =
  \frac{-3c_{7}g}{64M^{4}_{0}} \left \{\tilde{{\bf \Delta}}^{(2)},
  \sigma \cdot \left ( \tilde{{\bf \Delta}} \times \tilde{{\bf E}} - \tilde{{\bf E}}
  \times \tilde{{\bf \Delta}} \right ) \right \} \nonumber }\\
  \lefteqn{ -\frac{c_{8}g}{8M^{3}_{0}} \left \{ \tilde{{\bf \Delta}}^{(2)},
  \sigma \cdot \tilde{{\bf B}} \right \}
  - \frac{ic_{9}g^{2}}{8M^{3}_{0}} {\bf \sigma} \cdot \tilde{{\bf
  E}} \times \tilde{{\bf E}} \, .}
\eea
A \emph{tilde} signifies the use of improved versions of the
lattice operators \cite{TrottierQuarkonium}, and all operators are
tadpole-improved. Much evidence suggests that the \emph{Landau}
definition of the tadpole factor is superior to the
\emph{plaquette} definition \cite{TrottierQuarkonium,Landau}. We
have used both in our simulations.

The gluon action used for the quenched simulations is tadpole and
`rectangle' improved \cite{LepageImproved},
\be
\label{ImpGaugeAction} S_{G} = - \beta \sum_{n,\mu>\nu} \left (
\frac{5P_{\mu\nu}}{3u_{0}^{4}} - \frac{(R_{\mu\nu} +
R_{\nu\mu})}{12u_{0}^{6}}  \right ) \, ,
\ee
where $P_{\mu\nu}(n)$
and $R_{\mu\nu}(n)$ are the traces of $1 \times 1$ plaquettes and
$2 \times 1$ rectangles of link operators respectively. We used
smeared operators for the $^{1}S_{0}$, $^{3}S_{1}$, $^{1}P_{1}$,
$^{3}P_{0}$, $^{3}P_{1}$ and $^{3}P_{2}$ states, previously
detailed in \cite{TrottierQuarkonium}.

\section{Simulation Details}

We performed simulations with the NRQCD Hamiltonian above,
truncated to \ovfour and \ovsix, with both the Landau and
plaquette tadpole factor. The MILC collaboration generously
provided us with an ensemble of 200 unquenched gauge field
configurations of volume $16^3 \times 32$, created with the Wilson
action at $\beta = 5.415$ and $n_f = 2$ staggered dynamical quarks
with $m = 0.025$ ($m_{ps}/m_{v} \simeq 0.45$).

We produced ensembles of 100 quenched configurations of volume
$12^{3} \times 24$, with both Landau and plaquette tadpole
factors. Using $\beta_L = 2.1$ and $\beta_P = 2.52$ gave roughly
the same lattice spacing as the unquenched configurations. As
heavy mesons are small, the difference in lattice volume between
the quenched and unquenched configurations should not effect our
results significantly. The lattice spacing was determined for each
ensemble using the spin-averaged $P$--$S$ splitting. The bare
charm mass $M_{0}$ was tuned by requiring that the kinetic mass of
the $^{1}S_{0}$ charmonium state, defined by $E({\bf p})-E(0) =
{\bf p}^2/2M_k$, agrees with the experimental mass of the
$\eta_{c}$. The simulations parameters are shown in Table
\ref{SimulationParameters}.

\begin{table*}[htb]
\renewcommand{\arraystretch}{1.15}
\caption{\label{SimulationParameters} Parameters used in
charmonium simulations.}
\begin{tabular*}{16cm}{c @{\extracolsep{\fill}} c c c c c} \hline
$\beta$ & $u_{0}^{P}$ & $u_{0}^{L}$ & $a$ (fm) & $aM_{0}$ & $s$
\\ \hline
\multicolumn{6}{c}{\emph{Quenched}} \\
2.52 & 0.874 & & 0.168(3) & 0.81 & 6 \\ 2.10 & & 0.829 & 0.181(3)&
1.15 & 4 \\ \hline
\multicolumn{6}{c}{\emph{Unquenched}} \\
5.415 & 0.854 & & 0.163(3) & 0.82 &  6 \\ 5.415 & & 0.800 &
0.163(3) & 1.15 &  4 \\ \hline
\end{tabular*}
\end{table*}

To increase statistics we calculated eight meson propagators on
each configuration (four spatial sources, two time slices).
Single-exponential fits were used for meson masses, while the
$S$-hyperfine splitting was extracted from a ratio fit.

\section{Results}

\begin{figure}[ht]
\begin{center}
  \scalebox{0.5}[0.5]{\includegraphics{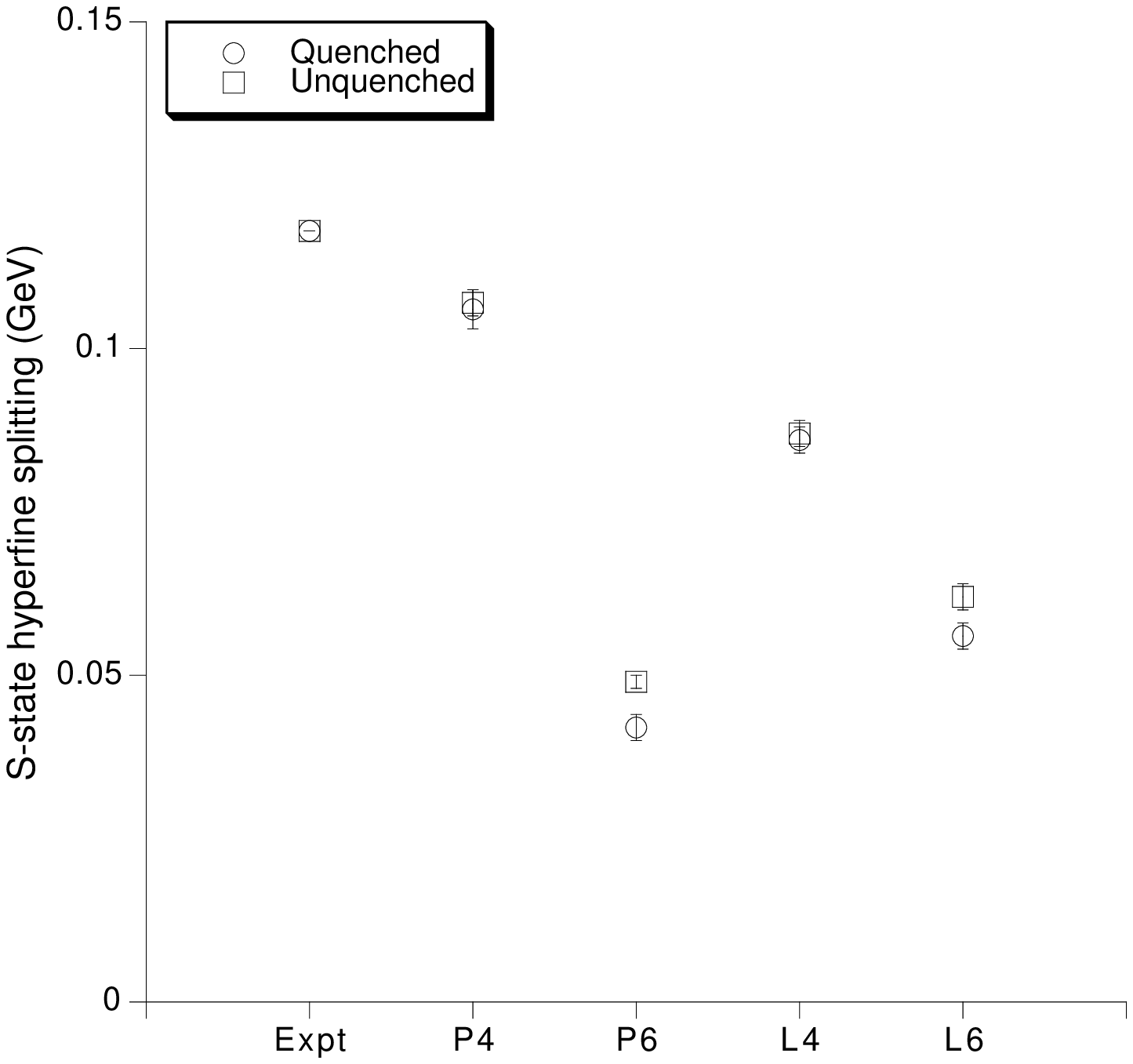}}
  \caption{\label{SHyp}$S$-hyperfine splitting.
   4, 6 = \ovfour, \ovsix; P, L = Plaquette, Landau tadpoles.}
\end{center}
\end{figure}

\begin{figure}[ht]
\begin{center}
\scalebox{0.5}[0.5]{\includegraphics{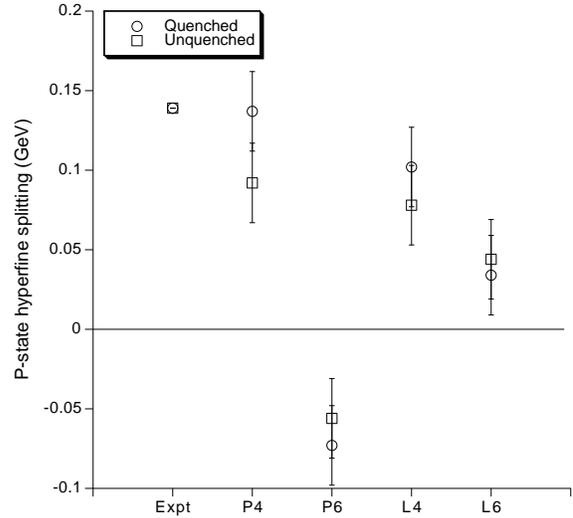}}
\caption{\label{PHyp}$P$-hyperfine splitting results.}
\end{center}
\end{figure}

Table \ref{Results} contains the final results (in GeV) for the
quenched and unquenched charmonium mass fits. Evidently, NRQCD
simulations of the charmonium system have a number of issues yet
to be resolved. This is most readily seen in the hyperfine
splittings, collected in Figures \ref{SHyp} and \ref{PHyp}.

The \ovsix corrections lead to a disturbingly large decrease in
the $S$-state hyperfine splittings, taking them as much as $60\%$
further away from the experimental values as first noted in
\cite{TrottierQuarkonium}. The plaquette-tadpole results are
strikingly bad, where the $^{3}P$ states appear in the wrong
order. This reversal is corrected in the Landau-tadpole
simulations, though the hyperfine splittings are still badly
underestimated.

The large discrepancies in spin-dependent splittings would be less
worrisome if quenching were seen to have a considerable effect on
the spectrum, as suggested in \cite{charm92}. Sadly, this does not
seem to be the case. There is some evidence for a $\sim 10 \%$
quenching effect in the \ovsix $S$-hyperfine data, though given
the apparent size of other systematic uncertainties no great
significance can be attached to these differences.

\begin{table*}[htb]
\caption {\label{Results} Quenched (top) and unquenched (bottom)
charmonium masses in GeV.}
\renewcommand{\arraystretch}{1.15}
\begin{tabular*}{16cm}{c @{\extracolsep{\fill}} c c c c c} \hline
State  & \multicolumn{2}{c}{$u_{0}^{P}$} &
\multicolumn{2}{c}{$u_{0}^{L}$} & Expt \\
  & ${\cal O}(v^{4})$ & ${\cal O}(v^{6})$
  & ${\cal O}(v^{4})$ & ${\cal O}(v^{6})$ & \\ \hline
$^{3}S_{1}$ & 3.086(2) & 3.022(2) & 3.066(2) & 3.036(1) & 3.098 \\
$^{1}P_{1}$ & 3.517(17) & 3.470(17) & 3.499(18) & 3.479(17) &
3.530 \\ $^{3}P_{0}$ & 3.439(16) & 3.522(20) & 3.426(15) &
3.441(17) & 3.400 \\ $^{3}P_{1}$ & 3.486(17) & 3.488(17) &
3.483(18) & 3.478(18) & 3.520 \\ $^{3}P_{2}$ & 3.576(17) &
3.449(16) & 3.528(22) & 3.475(17) & 3.570 \\ $^{3}S_{1}
-$$^{1}S_{0}$ & 0.106(3) & 0.042(2) & 0.086(2) & 0.056(2) & 0.118
\\ \hline
\end{tabular*}
\begin{tabular*}{16cm}{c @{\extracolsep{\fill}} c c c c c}
$^{3}S_{1}$ & 3.087(2) & 3.028(2) & 3.068(2) &
3.043(1) & 3.098 \\ $^{1}P_{1}$ & 3.514(17) & 3.475(17) &
3.500(17) & 3.486(16) & 3.530 \\ $^{3}P_{0}$ & 3.456(15) &
3.518(20) & 3.437(15) & 3.445(15) & 3.400 \\ $^{3}P_{1}$ &
3.501(17) & 3.499(17) & 3.490(17) & 3.479(17) & 3.520 \\
$^{3}P_{2}$ & 3.548(21) & 3.462(16) & 3.515(21) & 3.489(17) &
3.570 \\ $^{3}S_{1} -$$^{1}S_{0}$ & 0.107(2) & 0.049(1) & 0.087(2)
& 0.062(1) & 0.118
\\ \hline
\end{tabular*}
\end{table*}

\section{Other Systematic Errors}

These results suggest quenching errors alone will not account for
the discrepancies in the spectrum. We have seen, as others have
previously, large differences between the $u_{0}^{L}$ and
$u_{0}^{P}$ results. This is not surprising, as terms in the NRQCD
Hamiltonian linear in {\bf E} or {\bf B} differ by as much as
$30\%$ between the different tadpole schemes in our simulations,
an effect at least as important as quenching effects.

\subsection{Radiative Corrections}

We expect some effect on the spectrum from high-momentum modes
that are cut off by the finite lattice spacing. These effects
appear as \oas corrections to the coefficients $c_i$ in the NRQCD
Hamiltonian. To date, corrections have been found for just three
\ovfour terms: the $c_1$ and $c_5$ 'kinetic' terms
\cite{MorningstarPertNRQCD} and the spin-dependent $c_4$ term
\cite{TrottierHighBeta}. The corrections are $\sim 10 \% $ for the
bottom quark, but rise dramatically as the bare quark mass falls
below one (in lattice units).

To roughly estimate the effects of \oas corrections to \emph{all}
coefficients, we replaced the $c_{i} = 1$ with $c_{i} = 1 \pm
\alpha_{s}$. On our lattices, $\alpha_{s}(\pi/a) \simeq
\frac{g^{2}}{4\pi} \sim 0.2$. We used the calculated values for
$c_1$, $c_4$ and $c_5$, and varied the remaining coefficients
between 0.8 and 1.2. This changed the $S$- and $P$-hyperfine
splittings by as much as $40 \%$---while this is only a crude
estimate, it is clear that the effects of radiative corrections
are easily as important as quenching effects for heavy-quark
systems. Accurate determinations of the remaining \oas corrections
are sorely needed.

\subsection{Improving the Evolution Equation}

Equation (\ref{EvolutionEquation}) contains better-than-linear
approximations to the exponential $e^{Ht}$ for the $H_{0}$ terms
but only a linear approximation for the $\delta H$ terms. As the
high-order corrections are large for charmonium, this
approximation is perhaps too severe. Lewis and Woloshyn made a
similar suggestion in their NRQCD study of $D$ mesons
\cite{RandyRichard}.

We examined this possibility for the \ovsix terms by using an
improved form for the evolution equation that incorporates a
`stabilisation' parameter for the correction terms, replacing
\be
(1-a\delta H) \to \left (1-\frac{a\delta H}{s_{\delta}} \right
)^{s_{\delta}} \, .
\ee
A simulation using this replacement, with $s_{\delta}=4$ and all
other parameters as for the previous Landau-tadpole quenched
simulations, showed that the $S$-hyperfine splitting increased by
as much as $20$ MeV. Statistical uncertainties in the
$P$-hyperfine splittings were large, though a similar increase
seems likely.

\section{Conclusions}

While NRQCD is considered quite unreliable for the charm quark,
the variety of relativistic quark actions also tend to
underestimate the charmonium $S$-hyperfine splitting by at least
$30$--$40 \%$. Some have suggested that much of the remaining
discrepancy is due to quenching errors. Our results indicate this
is unlikely to be the case---we find at most a $10 \%$ effect from
unquenching in the hyperfine splittings with both \ovfour and
\ovsix NRQCD.

In contrast, the choice of tadpole factor and evolution equation
each affect hyperfine splittings by $20$ MeV or more---the
instability of the evolution equation should be investigated
further. Further, \oas radiative corrections may shift the
spin-splittings by as much as $40 \%$. While this is a crude
estimate, such sizeable corrections could easily swamp the small
quenching error, and need to be carefully analysed.

\vspace{0.2cm} Thanks to Howard Trottier and Randy Lewis for
stimulating discussions.


\end{document}